\documentclass[10 pt, conference]{ieeeconf}
\usepackage{cite}
\usepackage{amsmath,amssymb,amsfonts}
\usepackage{algorithmic}
\usepackage{comment}
\usepackage{graphicx}
\usepackage[inkscapelatex=false]{svg}
\usepackage{booktabs}
\usepackage{textcomp}
\newtheorem{assumption}{Assumption}

\newtheorem{theorem}{Theorem}
\newtheorem{remark}{Remark}
\newtheorem{lemma}{Lemma}
\newtheorem{proposition}{Proposition}

\newtheorem{definition}{Definition}

\def\BibTeX{{\rm B\kern-.05em{\sc i\kern-.025em b}\kern-.08em
    T\kern-.1667em\lower.7ex\hbox{E}\kern-.125emX}}
\begin{document}
\title{Constraint Horizon in Model Predictive Control}
\author{Allan Andre Do Nascimento$^*$, Han Wang$^{*,\dagger}$, Antonis Papachristodoulou and Kostas Margellos
\thanks{Submission for review: $17^{th}$ of March of 2025. AAdN, AP and KM acknowledge funding support by MathWorks. AP was supported in part by UK’s Engineering, Physical Sciences Research Council projects EP/X017982/1 and EP/Y014073/1. 
For the purpose of Open Access, the authors have applied a CC BY public copyright licence to any Author Accepted Manuscript (AAM) version arising from this submission. }
\thanks{* Equal contribution, $\dagger$ corresponding author. All authors are with the Department of Engineering Science, University of Oxford,  Parks Road, Oxford OX1 3PJ, UK (email: \{allan.adn, han.wang, antonis, kostas.margellos\}@eng.ox.ac.uk).} }

\maketitle

\begin{abstract}
In this work, we propose a Model Predictive Control (MPC) formulation incorporating two distinct horizons: a prediction horizon and a constraint horizon. This approach enables a deeper understanding of how constraints influence key system properties such as suboptimality, without compromising recursive feasibility and constraint satisfaction. In this direction, our contributions are twofold. First, we provide a framework to estimate closed-loop optimality as a function of the number of enforced constraints. This is a generalization of existing results by considering partial constraint enforcement over the prediction horizon. Second, when adopting this general framework under the lens of safety-critical applications, our method improves conventional Control Barrier Function (CBF) based approaches. It mitigates myopic behaviour in Quadratic Programming (QP)-CBF schemes, and resolves compatibility issues between Control Lyapunov Function (CLF) and CBF constraints via the prediction horizon used in the optimization. We show the efficacy of the method via numerical simulations for a safety critical application.
\end{abstract}

\section{Introduction}
\label{sec:introduction}
Modelling and control design require a detailed description of feedback component properties, including actuator and sensor limits, which constrain system inputs and outputs. In safety-critical applications, additional constraints may apply. While these must be strictly enforced, performance should still be optimized. 
Model Predictive Control (MPC) \cite{kouvaritakis2016model} addresses this by enforcing constraints in the optimization problem while ensuring recursive feasibility. This is typically achieved via constraint enforcement over the full prediction horizon along with a terminal invariant set constraint. Optimality is attained with respect to the formulated cost function. MPC is an approximation of the corresponding infinite-horizon constrained optimal control problem, as such, it is crucial to assess its suboptimality.

\subsection{Related work}
Closed loop optimality of MPC has previously been analyzed in many different ways. One of the first works addressing this question was \cite{shamma1997linear}, in which for a linear discrete time system, a bound on closed-loop sub-optimality was calculated given that finite horizon optimal value functions were known. In \cite{grune2008infinite}, sub-optimality estimation of MPC was also addressed, this time for a nonlinear system by means of suitable upper bound estimations on the open loop value functions. This work also presents a condition for stability based on the optimality estimate. Many subsequent analyses building on this work followed. In \cite{grune2009practical}, the authors studied the effect of partially bounding the open-loop trajectory for each MPC problem on the upper bound estimated for closed-loop optimality. This result is derived for a system with the same type of constraints throughout the entire prediction horizon and assumes optimal input and state pair availability. Other papers, such as \cite{worthmann2011stability,grune2017nonlinear,grune2019dynamic}, use a varying parameter to bound the open-loop value function to running cost ratio, replacing the fixed parameter used in prior work. These papers provide a tighter suboptimality bound via varying parameters. Relevant results have been derived not only for stabilizing MPC but also for economic MPC as in \cite{grune2013economic,grune2019dynamic,faulwasser2018economic,kohler2023stability}.

When it comes to constraint satisfaction and recursive feasibility, not only imposing invariance constraints at the end of the prediction have been attempted. For instance, in \cite{mhaskar2006stabilization} a Lyapunov-based MPC is proposed, where the first step in the prediction horizon is constrained with a Control Lyapunov Constraint. In \cite{de2008lyapunov}, a Lyapunov-based MPC is studied to explicitly characterize the stability region based on the number of data sample losses in a networked system. This is performed via a Lyapunov decaying condition being enforced over an arbitrary part of the prediction horizon.

A similar trend recently emerged using Control Barrier Functions (CBFs) \cite{prajna2004safety}, \cite{wieland2007constructive}, initially proposed as a way to address the short-sightedness of Quadratic Programming CBF filters, applied to safety-critical systems \cite{ames2014control}. In \cite{wabersich2021predictive} for instance, an MPC-like approach is investigated to derive a predictive safety filter. One of the first works designing safe controllers by explicitly using MPC with CBF constraints is \cite{zeng2021safety}, which highlights the benefits of CBFs over distance constraints through simulation. Other works using CBF as MPC constraints  are: \cite{do2023game}, in which safety is sought for distributed systems and in \cite{do2024probabilistically}, where probabilistic safety is obtained. Unlike \cite{zeng2021safety}, these works enforce safety constraints only in the first prediction step, leading to short-sighted long-term safety guarantees.

\subsection{Main contributions}

In this work, we propose an MPC formulation with two different horizons—prediction and constraint—following \cite{de2008lyapunov}, which do not coincide. This approach examines the impact of varying constraints while ensuring recursive feasibility and constraint satisfaction.

This approach contributes in two key directions. First, we propose a method to estimate closed-loop optimality based on varying constraints, generalizing prior results. Unlike \cite{grune2008infinite, grune2009practical}, our bounds consider constraints that do not apply over the entire prediction horizon, recovering results in \cite{grune2008infinite} when applied to the full horizon.

Second, we enhance safety-critical control by offering a general framework that adopts an MPC setting and imposes safety constraints at some steps over the prediction horizon. Compared to QP-CBF schemes, it mitigates myopic behaviour, reducing aggressive actions and abrupt collision avoidance. It also provides stability guarantees and resolves CLF-CBF compatibility issues via receding horizon control and appropriate cost functions.

Finally, by integrating MPC and CBFs, we provide the first rigorous analysis quantifying closed-loop optimality relative to varying constraint enforcement. This provides an in-depth analysis and algorithm design guidance if compared to \cite{do2023game,zeng2021safety,do2024probabilistically}, while generalizing their problem formulation.

This paper is structured as follows: Section \ref{sec:formulation} defines the problem, Section \ref{sec:mainres} presents optimality bounds and stability as a function of constraint horizons, Section \ref{sec:applications} covers safety-critical applications and simulations, and Section \ref{sec:conclusion} concludes the paper. 

\section{Constraint horizon in MPC - Problem formulation}\label{sec:formulation}

\begin{figure}[t!]
    \centerline{\includegraphics[width=0.95\columnwidth]{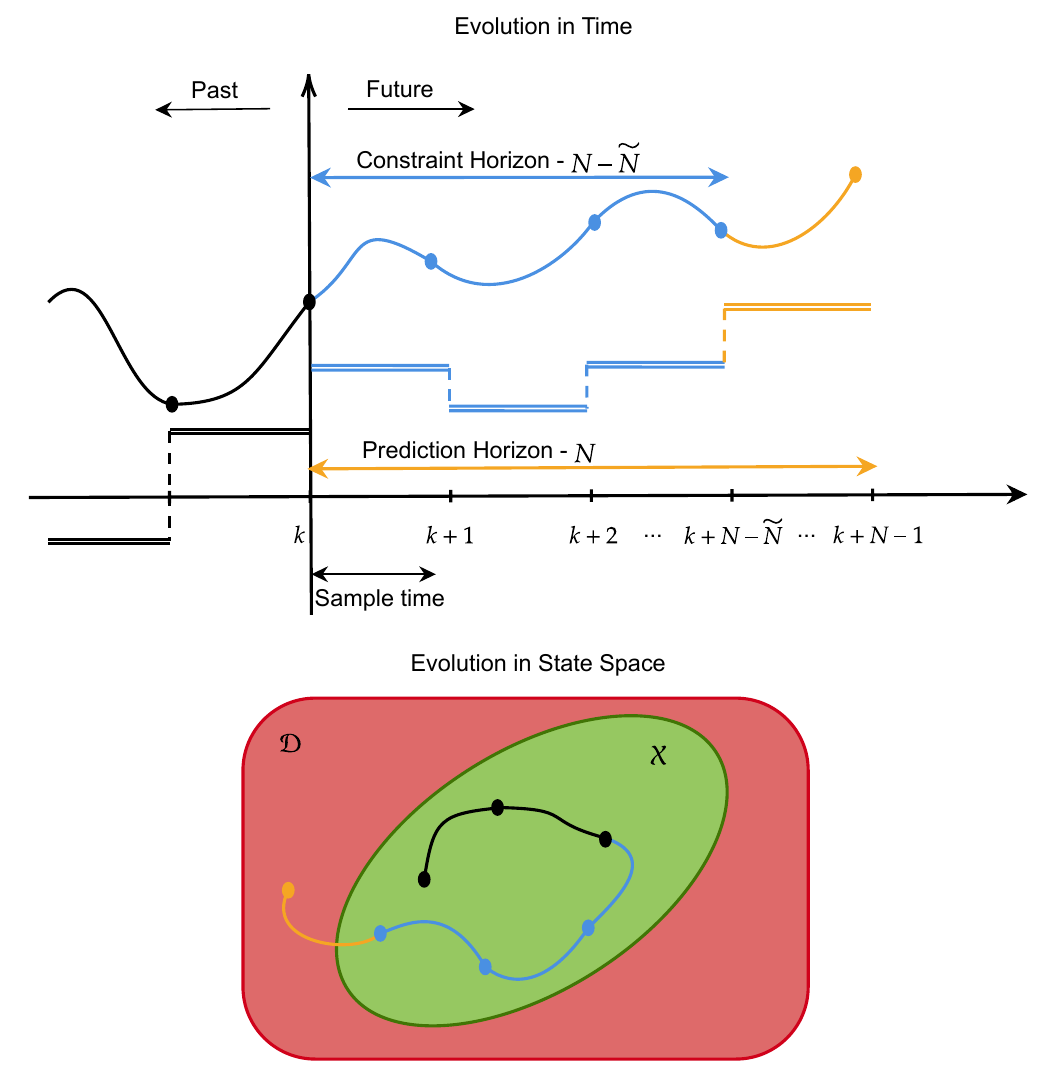}}
    \caption{Distinct horizons on MPC. Line coding: Single lines - system trajectories, double lines - system inputs. Color coding: Black - past actions, Blue - safe actions (within constraint horizon), Yellow - possibly unsafe actions (within prediction horizon but no constraints are enforced), Red - unsafe set, Green - safe set.}
    \label{fig:mpc}
\end{figure}

Consider compact sets $\mathcal{D}\subset\mathbb{R}^n$, $\mathcal{U}\subset\mathbb{R}^m$ and the discrete-time dynamical system
\begin{equation}\label{eq:system}
    x(k+1)=f(x(k),u(k)),
\end{equation}
where $x(k) \in \mathcal{D}$, $u(k) \in \mathcal{U}$ and $f(0,0)=0$. We consider an MPC problem at time $k$ with state $x(k)$, where constraints are not enforced over the entire prediction horizon $N$, but only over a shorter period termed constraint horizon. The final $\tilde N$ steps are unconstrained. We will sometimes refer to such a problem as partially constrained.
\begin{subequations}\label{eq:our-filter-new-new}
    \begin{align}
    &V_{N}^{\tilde N}(x(k)):=\min_{\{u_{i|k}\}_{i=0}^N}\quad \label{cost} \sum_{i=0}^{N-1}l(x(i|k),u(i|k)) \\
    &\mathrm{subject~to} \nonumber\\
    & \label{dyn_constr} x(i+1|k)=f(x(i|k),u(i|k)),i=0,\ldots,N-1, \\
    \label{init_state}
    &x(0|k)=x(k),  \\
    \label{inp_constr}
    &u(i|k)\in\mathcal{U},i=0,\ldots,N-1, \\
    \label{cbf_constr}
    & x(i+1|k) \in \mathcal{X}, i=0,\ldots,N-\tilde N. 
\end{align}
\end{subequations}
In \eqref{eq:our-filter-new-new} we use the notation $x(i|k), u(i|k)$ to denote the state and input, $i$ steps ahead of the current time step $k$. We restrict our analysis to stabilizing MPC problems, which implies that the cost $l(\cdot,\cdot)$ is positive definite, where $l(0,0)=0$. Finally $\mathcal{U}$ defines the input constraints and $\mathcal{X} \subseteq \mathcal{D}$ the state constraints, in which we assume $\mathcal{X}$ to be a \emph{control invariant set} \cite[Definition 2.3]{blanchini1999set} and $ x(k) \in \mathcal{X}$. The difference between a traditional MPC formulation and \eqref{eq:our-filter-new-new} lies in \eqref{cbf_constr}. Here, state constraints act only during the \emph{constraint horizon} of length $N-\tilde{N}$, instead of acting over the \emph{prediction horizon} of length $N$, as is typically done. Figure \ref{fig:mpc} illustrates this concept.

\begin{remark}
In most of the MPC literature \cite{grune2008infinite}, $\mathcal{X}$ is not required to be invariant and is applied throughout the whole prediction horizon. Nonetheless, the minimum with respect to $u \in \mathcal{U}$ is always assumed to be attained. In \eqref{eq:our-filter-new-new} the feasible input set will be $\mathcal{U} \cap \mathcal{U}(x)$ where $\mathcal{U}(x) = \{u(i|k) \in \mathcal{U}, \: \textrm{s.t.}, \: x(i+1|k) \in \mathcal{X}\}, \:  i=0,\ldots,N-\tilde N$,  is the set of inputs satisfying \eqref{cbf_constr}. Since the minimum in $u$ cannot be attained if $\mathcal{U}\cap \mathcal{U}(x)=\varnothing$, we restrict ourselves to  control invariant sets $\mathcal{X}$. 
\end{remark}
\begin{remark}
The assumption on the invariance of $\mathcal{X}$ can be, at a first glance, perceived as conservative since, when comparing to traditional MPC formulations, we are demanding the system to start \emph{within} an invariant set, instead of entering a terminal invariant set by the end of the prediction horizon. Nonetheless, this new framework opens up different perspectives. First, it allows us to guarantee recursive feasibility by using constraint \eqref{cbf_constr} in \eqref{eq:our-filter-new-new} only for $i=0$. The possibility of partially constraining the system without jeopardizing recursive feasibility, which is not possible in the traditional setting, allows us to quantitatively evaluate the impact of the number of constraints in the closed-loop optimality estimation. Finally, invariance in the beginning of the prediction horizon may be a ``natural specification" depending on the application as we will see in the applications section.         
\end{remark}

Let $u^*(0|k),\ldots,u^*(N-1|k)$ be the optimal solution, also termed \emph{open-loop optimal control sequence\footnote{The optimal solution is implicitly a function of the initial state $x(k)$.}} of problem \eqref{eq:our-filter-new-new}. For notational purposes, let

\begin{equation}\label{eq:closed-loop-control}
    \mu_{p}(x):=u^*(N-p|k)(x),p\in\{1,\ldots,N\}.
\end{equation}
The dependence of $\mu_{p}(x)$ on time $k$ will be reflected by the state $x(k)$. 

We now derive a sequence of value functions $V_n^{\tilde N},n\in\{0,\ldots,N\}$ which will be important for our optimality proof later. For $n\in\{1,\ldots,\tilde N\}$, and assuming $x(k) \in \mathcal{X}$, the optimal control problem is formulated as
\begin{subequations}\label{eq:value-func-1}
    \begin{align}
    &V_{n}^{\tilde N}(x):=\min_{u(i|k)}\quad \sum_{i=N-n}^{N-1}l(x(i|k),u(i|k))\\
    &\mathrm{subject~to} \nonumber\\
    & x(i+1|k)=f(x(i|k),u(i|k)),\\
    &x(N-n|k)=x(k), \\
    &u(i|k)\in\mathcal{U},i=N-n,\ldots,N-1,
\end{align}
\end{subequations}
For $n\in\{\tilde N+1,\ldots,N\}$, and again assuming $x(k) \in \mathcal{X}$ the optimal control problem is given by
\begin{subequations}\label{eq:value-func-2}
    \begin{align}
    &V_{n}^{\tilde N}(x):=\min_{u(i|k)}\quad  \sum_{i=N-n}^{N-1}l(x(i|k),u(i|k))\\
    &\mathrm{subject~to} \nonumber\\
    &x(i+1|k)=f(x(i|k),u(i|k)),\\
    &x(N-n|k)=x(k), \\
    &u(i|k)\in\mathcal{U},i=N-n,\ldots,N-1,\\
    & x(i+1|k) \in \mathcal{X}, \:
    i=N-n,\ldots,N- \tilde{N}. \label{eq:inv-constr}
\end{align}
\end{subequations}
Specifically, when $n=N$, \eqref{eq:value-func-2} is the same as \eqref{eq:our-filter-new-new}. Given that there is no terminal cost in \eqref{eq:our-filter-new-new}, for $n=0$ the value function
\begin{equation}
    V_0^{\tilde N}=0.
\end{equation}

Here, the optimal control formulation for the value function $V_n^{\tilde N}(x)$ has two distinct forms, where the control invariant set constraints \eqref{eq:inv-constr} are incorporated in \eqref{eq:value-func-2}, but do not appear  in \eqref{eq:value-func-1}. This is a result of studying the effect of a varying constraint horizon in the optimal solution, via dynamic programming. Furthermore, notice that \eqref{eq:value-func-2} has been defined in the above format because the value functions $\{V_n^{\tilde N}(x)\}_{n=0}^N$ are defined backwards. Such a formulation is seamlessly amenable to the principle of optimality, based on which the optimal control sequences are defined. For $n\in\{1,\ldots,\tilde N\}$, $\{\mu_n(x)\}_{n=1}^{\tilde N}$ and value functions $\{V_{n}^{\tilde N}(x)\}_{n=1}^{\tilde N}$ satisfy
\begin{subequations}
    \begin{align} \label{eq:subeq_bpo1}
        &\mu_n(x)=\mathop{\arg\min}_{u\in\mathcal{U}}V_{n-1}^{\tilde N}(f(x,u))+l(x,u),\\
        &V_n^{\tilde N}(x)=V_{n-1}^{\tilde N}(f(x,\mu_n(x)))+l(x,\mu_n(x)).
    \end{align}
\end{subequations}
For $n\in\{\tilde N+1,\ldots,N\}$, the control invariant set constraint should be taken into account. Then, the optimal control sequence $\{\mu_n(x)\}_{n=\tilde N+1}^{N}$ and value functions $\{V_n^{\tilde N}(x)\}_{n=\tilde N+1}^N$ satisfy
\begin{subequations}
    \begin{align} \label{eq:subeq_bpo2}
        &\mu_n(x)=\mathop{\arg\min}_{u\in\mathcal{U}\cap\mathcal{U}(x)}V_{n-1}^{\tilde N}(f(x,u))+l(x,u), \\
        &V_{n}^{\tilde N}(x)=V_{n-1}^{\tilde N}(f(x,\mu_n(x)))+l(x,\mu_n(x)) \label{eq:value-iteration}.
    \end{align}
\end{subequations}

Note that in \eqref{eq:subeq_bpo2}, the minimization is over $\mathcal{U}\cap\mathcal{U}(x)$ as opposed to \eqref{eq:subeq_bpo1}, which is just over $\mathcal{U}$.
Moreover, in case $V^{\tilde N}_N(x(k))$ is known, one could also directly use the Bellman's principle of optimality in \eqref{eq:subeq_bpo1} and \eqref{eq:subeq_bpo2} to obtain \eqref{eq:closed-loop-control}. The \emph{closed-loop} controller is defined to be $\mu_N(x)$. In other words, the \emph{closed-loop} sequence is $\mu_N(x(0)), \ldots, \mu_N(x(k))$ which is applied to system \eqref{eq:system}, produces the closed-loop system dynamics
\begin{equation}\label{eq:closed-loop-dynamics}
    x(k+1) = f(x(k),\mu_N(x(k))).
\end{equation}

For the sake of brevity, we drop the subscript $\mu_N$ and the argument $k$ from $x$ in the sequel whenever no ambiguity arises.
At the next time iteration $k+1$, the optimal control problem \eqref{eq:our-filter-new-new} is solved again, this time for an initial state $x(k+1)$. This process is repeated for each new time step and state available. The associated \emph{closed-loop infinite horizon cost} of \eqref{eq:our-filter-new-new} is defined by
\begin{equation}\label{eq:closed-loop-cost}
    J_{\infty}^{N,\tilde N}(x)=\sum_{k=0}^\infty l(x(k),\mu_N(x(k))).
\end{equation}
In this work, our goal is to investigate the closed-loop optimal cost $J_\infty^{N,\tilde N}(x)$ of the MPC problem \eqref{eq:our-filter-new-new} and explore the impact of the \emph{prediction horizon} $N$ and the \emph{constraint horizon} $N-\tilde N$ on it. This optimality analysis will both provide us bounds and serve as the backbone of the stability analysis. 

\section{Main results} \label{sec:mainres}
\subsection{Optimality estimation}
Here we start by providing some definitions to give a concise answer to our previous question. The following lemma establishes the relationship between $J_{\infty}^{N,\tilde N}(x)$ and $V_N^{\tilde N}(x)$.
\begin{lemma} \cite[Proposition 2.2]{grune2008infinite}\label{lem:closed-optimality}
    Consider $N\ge 2$, $1\le \tilde N\le N-1$. Assume that
    \begin{equation}
        V_N^{\tilde N}(f(x,\mu_N(x)))-V_{N-1}^{\tilde N}(f(x,\mu_N(x)))\le (1-\alpha)l(x,\mu_N(x))
    \end{equation}
    holds for some $\alpha\in[0,1]$ and all $x\in\mathcal{X}$. Then $V_N^{\tilde N}(x)$ satisfies
    \begin{equation} \label{eq:dyn_prog}
        V_N^{\tilde N}(x)\ge V_N^{\tilde N}(f(x,\mu_N(x)))+\alpha l(x,\mu_N(x)),
    \end{equation}
    and
    \begin{equation}\label{eq:alpha}
        \alpha J_{\infty}^{N,\tilde N}(x)\le V_N^{\tilde N}(x),
    \end{equation}
    for all $x\in\mathcal{X}$.
\end{lemma}
Equation \eqref{eq:dyn_prog} is well known as the relaxed dynamic programming formulation. We briefly sketch the main idea for Lemma \ref{lem:closed-optimality}. Equation \eqref{eq:alpha} is obtained from \eqref{eq:dyn_prog} by rewriting it as:
\begin{equation} \label{eq:int_bd}
    V_N^{\tilde N}(x(k)) - V_N^{\tilde N}(x(k+1)) \ge\alpha l(x(k),\mu_N(x(k)))
\end{equation}
By summing \eqref{eq:int_bd} over $M$ sequential discrete time instances, we get:
\begin{align} \label{eq:sum_bd}
    & V_N^{\tilde N}(x(k)) \ge V_N^{\tilde N}(x(k)) - V_N^{\tilde N}(x(k+1+M)) \nonumber \\ 
    & \ge \alpha \sum^M_{j=0}l(x(k+j),\mu_N(x(k+j))) = \alpha J^{N,\tilde{N}}_M (x(k))
\end{align}
which by letting $M \rightarrow \infty$ and omitting the initial state time argument $k$, leads to \eqref{eq:alpha}.
The closed loop MPC performance is then upper-bounded by the open loop cost  $V_N^{\tilde N}(x)$ scaled by the inverse of $\alpha \in [0,1]$. 

We now characterize $\alpha$ as an explicit expression of $N$ and $\tilde{N}$. First, consider the assumption below:
\begin{assumption}\label{ass:value-cost}
    Consider problem \eqref{eq:our-filter-new-new}. For a given $N\in\mathbb{N}$ and $0\le \tilde N\le N-1$, $\tilde N\in\mathbb{N}$, there exists $\beta > 0$ such that the inequalities
    \begin{subequations}
        \begin{align}
        &V_{\tilde N+1}^{\tilde N}(x)\le (\beta+1)V_{\tilde N}^{\tilde N}(x) \label{eq:ass1eq1}\\
            &V_n^{\tilde N}(x)\le (\beta+1)l(x,\mu_n(x)),n\in\{\tilde N+1,\ldots,N\}\label{eq:ass1eq2}
        \end{align}
    \end{subequations}
    hold for all $x\in\mathcal{X}$.
\end{assumption}
On Assumption \ref{ass:value-cost}, both \eqref{eq:ass1eq1} and \eqref{eq:ass1eq2} need to be valid for $n=\tilde{N}+1$. We emphasize that both conditions can be fulfilled simultaneously. For instance, assume the existence of $\beta_1>0$ and $\beta_2>0$ satisfying \eqref{eq:ass1eq1} and \eqref{eq:ass1eq2} respectively. By choosing $\beta=\max\{\beta_1,\beta_2\}$,  \eqref{eq:ass1eq1} and  \eqref{eq:ass1eq2} will be fulfilled for all $n \geq \tilde{N}+1$.
Under this assumption, we state the main theorem:
\begin{theorem}\label{th:optimality}
    Consider problem \eqref{eq:our-filter-new-new} with $N\ge2$, $1\le \tilde N\le N-1$, $\beta>0$ such that $(\beta+1)^{N-\tilde N-1}>\beta^{N-\tilde N+1}$. Let Assumption \ref{ass:value-cost} hold. Then, inequality \eqref{eq:alpha} holds for all $x\in\mathcal{X}$ with 
    \begin{equation} \label{eq:optim_alpha}
    \alpha = 1-\frac{\beta^{N-\tilde N+1}}{(\beta+1)^{N-\tilde N-1}}.
\end{equation}
\end{theorem}
\vspace{2mm}

Before we provide the proof for Theorem \ref{th:optimality}, we first briefly discuss the derived bound.
Theorem \ref{th:optimality} estimates the closed-loop optimality of MPC problem \eqref{eq:our-filter-new-new} via the open-loop cost $V_N^{\tilde N}(x)$ and $\alpha$, which depends explicitly on the prediction and constraint horizons, $N$ and $N-\tilde{N}$. The value $\beta$ is obtained by fulfillment of conditions \eqref{eq:ass1eq1} and \eqref{eq:ass1eq2} for all $x \in \mathcal{X}$. One could make use of the system structure, via cost controllability \cite{grune2017nonlinear} for instance, to obtain an  \emph{offline} calculation of $\beta$. For now nonetheless, we compute $\beta$ \emph{online}. An example on how to do it will be presented in the next section.

Notably, if $\beta$ is fixed for any $N$ and $\tilde N$, then increasing $N$ while keeping $\tilde N$ fixed makes $J_{\infty}^{N,\tilde N}(x)$ closer to $V_N^{\tilde N}(x)$. Similarly, for fixed $N$, reducing $\tilde N$ (i.e., extending the constraint horizon) has the same effect. Thus, a longer prediction and/or constraint horizon helps achieve a closed-loop upper bound closer to $V_N^{\tilde N}$ by driving $\alpha \rightarrow 1$. These aspects will be explored in the numerical simulations. 

\subsection{Proof of Optimality bound \& stability considerations}
We now construct the proof for Theorem \ref{th:optimality}. Using Assumption \ref{ass:value-cost}, the following proposition establishes a preliminary result for the closed-loop optimality analysis.

\begin{lemma}\label{lem:value-func}
    Consider problem \eqref{eq:our-filter-new-new} with $N\ge2$, $1\le \tilde N\le N-1$. Let Assumption \ref{ass:value-cost} hold. Then, the inequality
    \begin{equation}\label{eq:lemma1eq1}
        \frac{(\beta+1)^{N-\tilde N-1}}{(\beta+1)^{N-\tilde N-1}+\beta^{N-\tilde N}}V_N^{\tilde N}\le V_{N-1}^{\tilde N}
    \end{equation}
    holds for all $x\in\mathcal{X}$.
\end{lemma}
\begin{proof}
    The proof follows similar procedure as that in \cite{grune2008infinite}. We first show that Assumption \ref{ass:value-cost} implies the estimate
    \begin{equation}\label{eq:lem1eq2}
        V_{n-1}^{\tilde N}(f(x,\mu_n(x)))\le \beta l(x,\mu_n(x)).
    \end{equation}
for all $n\in\{\tilde N+1,\ldots,N\}$, $x\in\mathcal{X}$. Using Assumption \ref{ass:value-cost} and the value iteration \eqref{eq:value-iteration}, we obtain
\begin{align}
    &V_{n-1}^{\tilde N}(f(x,\mu_n(x)))=V_n^{\tilde N}(x)-l(x,\mu_n(x))\nonumber\\
    \le ~&(\beta+1)l(x,\mu_n(x))-l(x,\mu_n(x))=\beta l(x,\mu_n(x)) 
\end{align}
for any $n\in\{\tilde N+1,\ldots,N\}$. Then, by induction over $n=\tilde N+1,\ldots,N$, we prove
\begin{equation}\label{eq:eta-V}
    \eta_n V_n^{\tilde N}(x)\le V_{n-1}^{\tilde N}(x),
\end{equation}
where 
\begin{equation}
    \eta_n=\frac{(\beta+1)^{n-\tilde N-1}}{(\beta+1)^{n-\tilde N-1}+\beta^{n-\tilde N}}.
\end{equation}
For the base case $n=\tilde N+1$, $\eta_n=\frac{1}{1+\beta}$, \eqref{eq:eta-V} coincides with the relation $V_{\tilde N+1}^{\tilde N}(x)\le (1+\beta)V_{\tilde N}^{\tilde N}$, which holds by Assumption \ref{ass:value-cost}.

For $n=\tilde N+1,\ldots,N$, the induction step $n\to n+1$ implies
\begin{align}
    V_n^{\tilde N}(x)&=V_{n-1}^{\tilde N}(f(x,\mu_n(x)))+l(x,\mu_n(x))\nonumber\\
    &\ge \left(1+\frac{1-\eta_n}{\beta+\eta_n}\right)V_{n-1}^{\tilde N}(f(x,\mu_n(x)))\nonumber\\
    &~~+\left(1-\beta\frac{1-\eta_n}{\beta+\eta_n}\right)l(x,\mu_n(x)) \nonumber\\
    &\ge \eta_n\left(1+\frac{1-\eta_n}{\beta+\eta_n}\right)V_{n}^{\tilde N}(f(x,\mu_n(x)))\nonumber\\
    &~~+\left(1-\beta\frac{1-\eta_n}{\beta+\eta_n}\right)l(x,\mu_n(x)) \nonumber\\
    &=\eta_n\frac{\beta+1}{\beta+\eta_n}\left(V_n^{\tilde N}(f(x,\mu_n(x)))+l(x,\mu_n(x))\right)\nonumber\\
    &=\eta_n\frac{\beta+1}{\beta+\eta_n}V_{n+1}^{\tilde N}(x),
\end{align}
where the first inequality follows from \eqref{eq:ass1eq2}, and the second one is due to the induction hypothesis in \eqref{eq:eta-V}. Moreover, 
\begin{align}
    \eta_n\frac{\beta+1}{\beta+\eta_n}&=\frac{(\beta+1)^{n-\tilde N}}{\beta(\beta+1)^{n-\tilde N-1}+\beta^{n-\tilde N +1}+(\beta+1)^{n-\tilde N-1}} \nonumber \\
    &=\frac{(\beta+1)^{n-\tilde N}}{(\beta+1)^{n-\tilde N}+\beta^{n-\tilde N+1}}=\eta_{n+1}
\end{align}
Thus, we conclude that $\eta_n V_n^{\tilde N}(x)\le V_{n-1}^{\tilde N}(x)$ for any $n\in\{\tilde N+1,\ldots,N\}$ and $x\in\mathcal{X}$. For $n=N$, we obtain \eqref{eq:lemma1eq1}.
\end{proof}

We are now ready to present the proof for Theorem \ref{th:optimality}.

\begin{proof}
    From Lemma \ref{lem:value-func} we obtain
    \begin{align*}
        V_{N}^{\tilde N}(x)-V_{N-1}^{\tilde N}(x)&\le \\
        & \left(\frac{(\beta+1)^{N-\tilde N-1}+\beta^{N-\tilde N}}{(\beta+1)^{N-\tilde N-1}}-1\right)V_{N-1}^{\tilde N}(x)\\
        &=\frac{\beta^{N-\tilde N}}{(\beta+1)^{N-\tilde N-1}}V_{N-1}^{\tilde N}(x)
    \end{align*}
Using this relation for $x=f(x,\mu_N(x))$, and substituting $V_{N-1}^{\tilde N}(f(x,\mu_N(x)))$ with $\beta l(x,\mu_N(x))$ using \eqref{eq:lem1eq2}, we obtain
\begin{align}
    &V_N^{\tilde N}(f(x,\mu_N(x)))-V_{N-1}^{\tilde N}(f(x,\mu_N(x)))\nonumber\\
    &\le \frac{\beta^{N-\tilde N+1}}{(\beta+1)^{N-\tilde N-1}}l(x,\mu_N(x)).
\end{align}
Therefore, we can use this inequality with Lemma \ref{lem:closed-optimality} to obtain the expression found for $\alpha$ in Theorem \ref{th:optimality} 
\end{proof}

The above results can also be used for a stability analysis as done in \cite{grune2008infinite}. For a positive definite function $l(\cdot,\cdot)$, $V^{\tilde N}_N (x)$ is a sum of positive definite functions, hence also positive definite. For $\alpha \in [0,1]$, which according to Theorem \ref{th:optimality} occurs when $(\beta+1)^{N-\tilde N-1}>\beta^{N-\tilde N+1}$, \eqref{eq:dyn_prog} can be rewritten as: 
    \begin{equation} \label{eq:Lyap} 
        - \alpha l(x,\mu_N(x)) \ge  V_N^{\tilde N}(x)(f(x,\mu_N(x))) - V_N^{\tilde N}(x).
    \end{equation}
In other words, since $V^{\tilde N}_N(x)$ is positive definite and respects \eqref{eq:Lyap}, $V^{\tilde N}_N(x)$ is a Lyapunov function. Thus we can ensure stability by enforcing $(\beta+1)^{N-\tilde N-1}>\beta^{N-\tilde N+1}$ when $\beta$ is fixed, via $N$ or $\tilde{N}$. For instance, we can derive a sufficient stability condition as a function of $N$ by also assuming a fixed $\tilde{N}$ and re-arranging the inequality to obtain:
\begin{equation}
    N \ge \left \lceil \tilde{N} + \frac{\log(\beta+1)+\log(\beta)}{\log{(\beta+1)}-\log{(\beta})} \right \rceil.
\end{equation} 

\section{Application: Safety Critical Control} \label{sec:applications}
Our methodology has been presented considering an arbitrary invariant set. Now, we apply the above formulation to safety critical systems.
\subsection{Control Barrier Functions}
A well-known technique to ensure invariance, mainly in the realm of safety, involves using Control Barrier Functions (CBFs). 
\begin{definition}\label{def:cbf}
    Consider again the compact set $\mathcal{D}\subset\mathbb{R}^n$, and system \eqref{eq:system}. Let $\gamma(\cdot)$ be an extended class-$\mathcal{K}$ function satisfying $0<\gamma(b(x)) \leq b(x)$. A $\mathcal{C}^0$ function $b(x):\mathcal{D}\to \mathbb{R}$ is called a Control Barrier Function (CBF) if
    \begin{enumerate}
        \item $\mathcal{B}:=\{x\in\mathcal{D}:b(x)\ge 0\}\subseteq\mathcal{X}$.\label{def:bullet1}
        \item  For any $x\in\mathcal{D}$, there exists $u(x) \in \mathcal{U}$, such that $b(f(x,u(x)))-b(x)\ge -\gamma(b(x))$.\label{def:bullet2}
    \end{enumerate}
\end{definition}

This provides a principled way to determine the invariant set in the framework presented in the previous sections.
From now on, let $\mathcal{X}$ in \eqref{cbf_constr} be defined by $\mathcal{B}$, where CBF constraints ensure system safety. This restriction offers many extra tangible benefits. If compared to traditional MPC safety enforcement such as distance constraints, CBFs provide an estimate of the problem's feasibility region, enhance robustness by enabling recovery to the safe set when the system initializes outside of it and, recursive feasibility is ensured even with a short prediction horizon. If compared to MPC-CBF methods such as \cite{zeng2021safety},\cite{do2023game}, and \cite{do2024probabilistically}, we generalize their formulation by allowing a tunable number of constraints while providing a closed-loop sub-optimality measure.

We illustrate this method by numerically evaluating the MPC CBF framework discussed here. 
\subsection{Numerical simulation}
We now evaluate the bounds given by Theorem \ref{th:optimality}. To this end, for problem \eqref{eq:our-filter-new-new}, we consider a linear double integrator system of the form $x(i+1|k) = Ax(i|k)+Bu(i|k)$, where the state $x(i|k)=[p_x(i|k),p_y(i|k),v_x(i|k),v_y(i|k)]^T$ collects the positions and velocities in the $x,y$ coordinates and the input $u(i|k)=[a_x(i|k), a_y(i|k)]^T$ contains the accelerations in the $x,y$ coordinates. We use a quadratic cost $l(x(i|k),u(i|k))=x(i|k)^TQx(i|k)+u(i|k)^TRu(i|k)$ and input constraints $\mathcal{U}=\{u(i|k) \in \mathbb{R}^2, s.t. \; -2\leq u_r(i|k) \leq 2\}$, where $r=\{1,2\}$ are the row entries of $u(i|k)$, and $i=0,\dots, N-1$ as in \eqref{dyn_constr} and \eqref{inp_constr}.

\begin{figure}[h]
\centering
\includegraphics[width=0.46\textwidth]{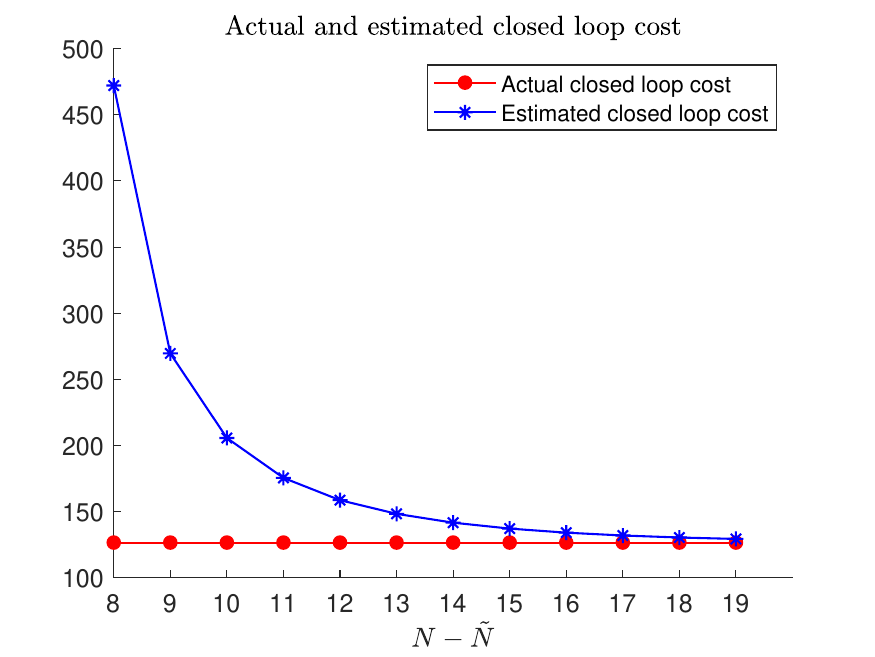}\\
\caption{\label{fig:cost_comp} 
Comparison between calculated and estimated closed loop costs according to $\eqref{eq:optim_alpha}$.
}
\end{figure}

As for the state constraint \eqref{cbf_constr} we will chose $|v_x(j|k)|+|v_y(j|k)| \leq 2$ for the velocity and for the positions, the CBF \emph{candidate functions}
\begin{subequations}
    \begin{align}
       & h_1(x(j|k)) = \frac{5}{9}p_x(j|k)+p_y(j|k)+\frac{0.5}{9}, \\
       & h_2(x(j|k)) = p_x(j|k)-p_y(j|k)+1.6,
    \end{align}
\end{subequations}
which restrain a region in $(p_x,p_y)$ where the system cannot enter. We then subject the system's positions to CBF constraints via $h_1(x(j+1|k))\geq(1-\gamma)h_1(x(j|k))$ and $h_2(x(j+1|k))\geq(1-\gamma)h_2(x(j|k))$, with $\gamma=0.8$ for $j=0,\dots,N-\tilde{N}$. To force the system to actively avoid the unsafe region we push it towards the unsafe region by setting its initial state to $x(0)=[-0.8,0.6,-0.45,0.65]$, and would like to control the system to the origin. In order to evaluate \eqref{eq:alpha}, we solve \eqref{eq:our-filter-new-new} repeatedly for $N=20$ and different values of $\tilde{N}$. For each value of $\tilde{N}$, we compute both $J^{N, \tilde{N}}_{\infty}(x_0)$ and $\beta$. We choose $\beta$ by testing conditions of Assumption \ref{ass:value-cost}. For each $N$, $\tilde{N}$ and $x(k)$ along the closed loop trajectory we compute according to conditions of Assumption \ref{ass:value-cost} intermediary values of $\beta$. By the end of each simulation with fixed values of $N$ and $\tilde{N}$, we generate a matrix $B_{\beta} \in \mathbb{R}^{{N-\tilde{N}+1}\times T}$, where $T$ is the total number of samples, and assign to $\beta$ the largest element of $B_{\beta}$, which will be used to calculate $\alpha$ as in \eqref{eq:optim_alpha}. We show results in Figure \ref{fig:cost_comp} for $N-\tilde{N}=8, \dots,19$.

Figure \ref{fig:cost_comp} indicates a reasonably tight bound that improves monotonically, though not for all $N, \tilde{N}$ with positive $\alpha$. However, a threshold likely exists where increasing $N-\tilde{N}$ ensures monotonic improvement, asymptotically reaching $\alpha \rightarrow 1$. While the bounds are tight here, we anticipate they may loosen if $\beta$ is fixed over the full safe set rather than the optimal trajectory, motivating a study of bounds with varying $\beta$.

For values of $N-\tilde{N}=\{1,\dots,7\}$ the assumption of $(\beta+1)^{N-\tilde N-1}>\beta^{N-\tilde N+1}$ in Theorem \ref{th:optimality} is not fulfilled. This results in $\alpha <0$, meaning the bound is not applicable. However, simulations remain feasible, ensuring a safe controller that drives the system to equilibrium. This highlights that $\alpha>0$ is a sufficient condition for stability. 

An interesting observation in numerical investigations was a non-increasing pattern in the actual closed loop cost with an increase in $N-\tilde{N}$. We display $J^{N,\tilde{N}}_{\infty}(x(0))$ for different values of $N$ and $N-\tilde{N}$ in the table below.

\begin{table}[htbp]
\centering
\textbf{Values of $J^{N,\tilde{N}}_{\infty}(x(0))$ for varying prediction and constraint horizon}
\vspace{10pt}
\addtolength{\tabcolsep}{-4pt}
\begin{tabular}{|c|c|c|c|c|c|}
\hline
     & $N-\tilde{N}=1$ & $N-\tilde{N}=3$ & $N-\tilde{N}=5$ & $N-\tilde{N}=7$ & $N-\tilde{N}=9$ \\ \hline
N=6 &  143.4820    &   126.5533   &   126.4878    &   N.A.    &  N.A.   \\ \hline
N=13 &    139.5191   &    126.4512   &   126.4298    &  126.4298    &   126.4298   \\ \hline
N=20 &   138.9538    &  126.4462     & 126.4295     &    126.4294   &    126.4294   \\ \hline
\end{tabular}
\label{tb:clcost}
\end{table}

\begin{figure}[h]
\centering
\includegraphics[width=0.46\textwidth]{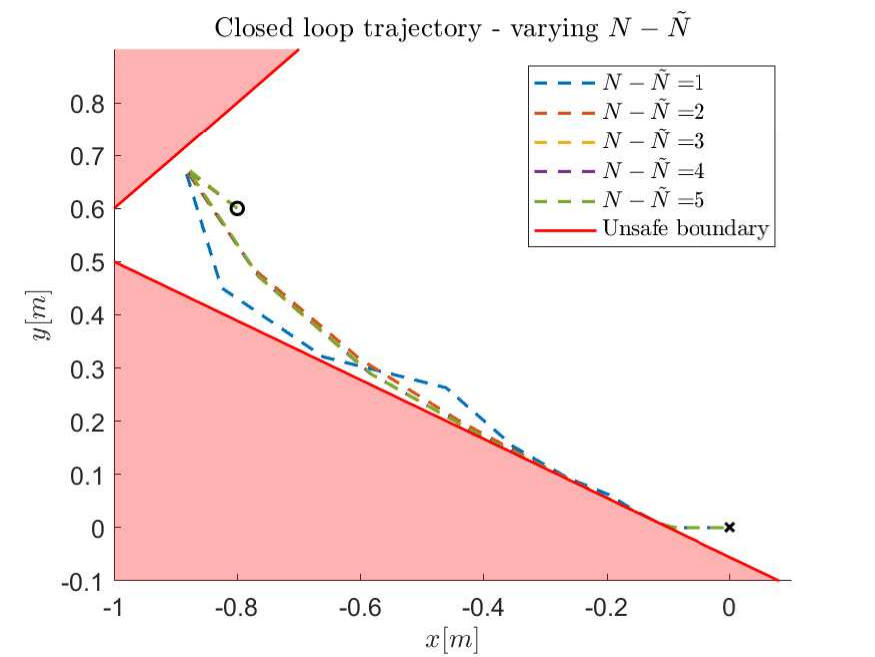}\\
\caption{\label{fig:traj_comp} 
System trajectories for $N=6$ and different values of $N-\tilde{N}$. Circle and cross are start and endpoint respectively.
}
\end{figure}

The closed-loop cost improvement was more pronounced for smaller $N-\tilde{N}$. The principle that a longer prediction horizon improves the closed-loop cost is also maintained. These suggest that increasing constraints may be beneficial, as delayed re-planning increases cost. Figure \ref{fig:traj_comp} illustrates this via different closed loop trajectories for $N=6$.

\section{Conclusion} \label{sec:conclusion}

In this work we have presented an MPC formulation possessing a prediction and a constraint horizon. We have shown, via sufficient conditions how this may affect closed loop bound optimality estimation and stability. We discussed this general framework under the lens of safety, and illustrated its behavior via simulation. This work opens up multiple future investigation directions, including how to estimate $\beta$ offline, how to adapt current results to a varying $\beta$ value and further investigation of the conjecture that more constraints yield lower closed loop cost.  

\section*{Acknowledgment}

Authors thank Professors Mark Cannon, Lars Grune and Karl Worthmann for their insightful suggestions kindly provided during discussions.

\bibliographystyle{IEEEtran}

\bibliography{refs.bib}

\end{document}